\documentclass[10pt,conference]{IEEEtran}
\IEEEoverridecommandlockouts

\usepackage{cite}
\usepackage{url}
\usepackage{booktabs}
\usepackage{tabularx}
\usepackage{array}
\usepackage{xspace}

\newcommand{\cisbs}{CISBs\xspace}
\newcommand{\bench}{CISB-Bench\xspace}

\begin{document}
\title{CISB-Bench: An Auditable Source--IR Dataset of Compiler-Introduced Security Bugs}
\author{\IEEEauthorblockN{Saatvik Pradhan, Anuroop Saini, Pranav Attrey, and Suman Saha}
\IEEEauthorblockA{The Pennsylvania State University\\
University Park, PA, USA\\
\{szp6118, abs7800, pka5235, szs339\}@psu.edu}}
\maketitle

\begin{abstract}
Compiler-introduced security bugs (CISBs) arise when an optimization, lowering,
or instrumentation decision changes a security-relevant property of the
generated program. They are difficult to study because their evidence is
distributed across issue reports, reduced tests, historical configurations, and
compiler artifacts; a security-related report also does not imply that every
associated reduction establishes a security-bearing compiler failure. We
present \bench, an auditable dataset of 429 exact C-program rows mined from
GCC and LLVM. Each row contains its C reduction, a standardized LLVM IR
analysis bundle at \texttt{-O0} through \texttt{-O3}, public provenance, a
final binary label, and a primary mechanism or boundary annotation. Two reviewers independently labeled the
fixed corpus, agreeing on 369 rows (86.0\%, Cohen's $\kappa=0.662$); the 60
disagreements were adjudicated. The final dataset comprises 280 CISBs and 149
hard non-CISB cases. The prediction task is to recover this reviewed exact-row
label from the supplied artifacts; it is not a claim that standardized IR alone
reproduces every historical compiler failure. We characterize the security mechanisms and evidence
boundaries represented by the corpus, and demonstrate how its paired artifacts
support source-only, IR-aware, and joint analyses. \bench provides a reusable,
inspectable target for compiler-security mining and detection research.
\end{abstract}

\begin{IEEEkeywords}
compiler security, mining software repositories, benchmark dataset, GCC, LLVM
\end{IEEEkeywords}

\section{Introduction}
\label{sec:introduction}

Compilers are part of the security boundary between a source program and the
executable that ultimately runs. Optimizations, target lowering, and security
instrumentation are expected to preserve relevant behavior, yet each can alter
a property on which a security mechanism relies. The resulting
correctness--security gap arises when a transformation meets a conventional
correctness criterion but defeats a security property~\cite{dsilva2015gap}.
Examples include optimizations that eliminate secret-erasure stores
\cite{yang2017dse}, sanitizer-eliding transformations
\cite{isemann2023dontlookub}, and compiler-created double fetches that weaken
control-flow integrity~\cite{xu2023warpattack}. Other work documents the
security consequences of compiler side effects and secret-dependent compiled
code~\cite{simon2018whatyouc,kaufmann2016constanttime,schneider2025breakingbad}.
We call such failures \emph{compiler-introduced security bugs} (\cisbs).

CISBs are challenging to mine and evaluate at scale. Their evidence is rarely
contained in a single artifact: an issue report may describe the affected
target, an old compiler version, a patch, and multiple reduced programs. A
reduction can be a complete witness, but it can also be a helper, a control, or
a fragment that no longer instantiates the reported mechanism. Conversely, a
compiler report can mention an overflow, crash, or optimization while
establishing only a warning problem, a compiler-internal failure, or an
ordinary wrong result. Security terminology and issue-family membership are
therefore insufficient ground truth for an exact program.

Compiler-testing methods discover semantic miscompilations
\cite{yang2011csmith,le2014emi}, empirical studies characterize GCC and LLVM
reports~\cite{sun2016compilerbugs,zhou2021optimization}, and prior security
work establishes that CISBs matter~\cite{xu2023silentbugs}. What remains
missing is an inspectable exact-row decision boundary with paired source/IR
evidence and difficult negatives.

We present \bench, 429 exact C reductions from GCC and LLVM, each with public
provenance and standardized LLVM IR at \texttt{-O0}--\texttt{-O3}. A
fixed-corpus, three-gate review produced 280 CISBs, 149 hard non-CISBs, and
mechanism or boundary annotations. The benchmark targets the reviewed exact-row
label from released C and IR; its standardized IR supports uniform analysis
rather than replaying every historical toolchain.

The distinguishing unit is an exact C reduction paired with standardized IR and
a reviewed compiler-to-security decision, rather than a report, generated test,
or source-level CVE instance (Table~\ref{tab:positioning}).

\begin{table}[t]
\caption{Positioning of \bench relative to representative resources. The
comparison concerns the released benchmark target, not the original value of
the other resources.}
\label{tab:positioning}
\centering
\footnotesize
{\setlength{\tabcolsep}{3pt}
\begin{tabularx}{\columnwidth}{p{0.29\columnwidth}X}
\toprule
Resource & Difference from \bench \\
\midrule
Csmith / EMI~\cite{yang2011csmith,le2014emi} & Generated tests and execution-based correctness evidence. \\
GCC/LLVM and CISB studies~\cite{sun2016compilerbugs,zhou2021optimization,xu2023silentbugs} & Report-level characterization or CISB study, not reviewed paired source--IR hard negatives. \\
Vulnerability datasets~\cite{lu2025icvul,ji2026rustxec,kahapola2026javabackports} & Different language, unit, or task. \\
\bench & Exact C rows, standardized IR, CISB target, hard negatives, and rationale. \\
\bottomrule
\end{tabularx}
}
\end{table}

\section{CISB Definition and Dataset Design}
\label{sec:definition}

\subsection{Exact-row definition}

We define a CISB as an exact program row for which three conditions hold.
First, \emph{compiler causation} (G1): an optimization, lowering decision, ABI
handling rule, or instrumentation action causes or materially enables the
failure. Second, \emph{semantic or protection change} (G2): that action changes
executable behavior or a requested runtime protection. The relevant change can
be a value, address, store, branch, layout, calling convention, or generated
check. Third, \emph{security relevance} (G3): the changed behavior affects a
supported security-bearing operation, including memory or concurrency safety,
integrity, confidentiality, availability, or a protection mechanism.

All three gates are required. Thus, a compiler crash does not create a
vulnerable executable; a missed warning is not a lost runtime protection; and
a changed numerical result is not a CISB without a supported security-bearing
consumer. We also distinguish source-authored unsafe or undefined behavior
from a separate compiler-created security mechanism. A policy-dependent class
retains cases in which compilation amplifies disclosed source undefined
behavior into a broader security-relevant manifestation; users applying a
strict defined-source policy can identify that class separately. This boundary
is motivated by prior work on optimization-unstable code and the security
consequences of undefined behavior~\cite{wang2012undefined,wang2013stack,lee2017taming}.

\subsection{Why exact rows and hard negatives}

The unit of analysis is a source reduction and its associated IR artifacts,
rather than a report title or issue family. This prevents a report's security
context from being inherited by every related test. It is especially important
because minimization can omit a dereference, target configuration, or consumer
that was essential to the original failure; reduction can also change
definedness assumptions~\cite{regehr2012reduction,lecoeur2023reconditioning}.
Upstream reports are retained as provenance and historical context, not as
automatically trusted labels. It also avoids treating issue-level duplicates or
closely related reductions as independent security witnesses
\cite{chen2013taming,yang2023dedup,marcozzi2019matter}.

The non-CISB class is consequently a designed boundary set, not random
background code. It contains diagnostic-only cases, compiler-process failures,
incomplete reductions, source-authored unsafe behavior, semantically benign
transformations, and ordinary wrong-code cases with no supported security
consequence. A method that merely recognizes words such as ``overflow,''
``sanitizer,'' or ``optimization'' should therefore produce false positives;
the benchmark evaluates prediction of the reviewed exact-row label, not
independent proof of the complete G1--G3 chain. Establishing the historical
mechanism may additionally require issue evidence and the affected compiler
configuration.
A non-CISB label means only that the retained evidence does not establish all
three gates for that exact row under this policy. It is not a general assurance
that the source program, the historical issue, or an execution in another
configuration is secure.

\section{Mining and Artifact Construction}
\label{sec:construction}

\subsection{Candidate retrieval}

We mined public GCC Bugzilla and LLVM issue-tracker evidence. Retrieval was
completed in July 2025; the resulting snapshot contains reports spanning many
earlier years. The archived construction snapshot retrieved 35,711 reports
(6,271 GCC and 29,440 LLVM).
We then retained 2,036 cases for which a self-contained C program and four
nonempty standardized LLVM IR artifacts could be assembled. Reports without a
retained C reduction or a complete artifact bundle were not eligible. This
criterion excludes many real reports, but it is necessary for a dataset whose
instances can be inspected and reused without a proprietary environment. This choice follows a
broader reproducibility tradition in mined software datasets, where an
available issue report alone is insufficient to make a case evaluable
\cite{tomassi2019bugswarm,bui2022vul4j,silva2024gitbugjava}.

Candidate selection used two independent triage signals to prioritize review.
First, a weighted keyword score over report text and tracker metadata used
security, memory-safety, optimization, control-flow, integer,
undefined-behavior, and compiler-component terms; a score cutoff of 10 retained
1,075 cases. Second, a fixed structured language-model screen using the OpenAI
model identifier \texttt{gpt-4.1-mini} read each report's summary, compiler
origin, component, severity, and tracker keywords; the dated backend snapshot
was not retained. The screen returned a candidate flag, confidence, proposed
impact, and vulnerability type. Retention depended only on a true candidate
flag, with no confidence cutoff, yielding 1,241 cases. Neither screen consumed
the released C/IR bundle or supplied a benchmark label.

We reviewed only their 429-case intersection. Requiring both a lexical signal
and an independent semantic screen concentrated finite manual review on cases
with two kinds of preliminary support, while preserving the entire resulting
candidate set for review rather than silently discarding later false positives.
Thus, the intersection is neither a prevalence sample nor a recall-oriented
census of compiler-security reports. Terms such as ``overflow'' and ``crash''
remain ambiguous; the later review determines the binary target. The archived
artifact contains the exact query groups, weights, structured-screen prompt,
and construction scripts.

\begin{table}[t]
\caption{Construction funnel. The fixed 429-row corpus is reviewed after
retrieval; no row is added or removed during final labeling.}
\label{tab:funnel}
\centering
\small
\begin{tabularx}{\columnwidth}{Xr}
\toprule
Stage & Cases \\
\midrule
Retrieved GCC/LLVM reports & 35,711 \\
Eligible C reductions with four IR artifacts & 2,036 \\
Keyword-triage candidates & 1,075 \\
Automated-triage candidates & 1,241 \\
Intersection / fixed review corpus & 429 \\
Final released rows & 429 \\
\bottomrule
\end{tabularx}
\end{table}

\subsection{Standardized source--IR analysis artifacts}

Each row includes a stable \texttt{sample\_id}, C source and hash, compiler and
issue provenance, final label, taxonomy/rationale, and Clang 20.1.2 LLVM IR at
\texttt{-O0}--\texttt{-O3} for
\texttt{arm64-apple-macosx15.0.0}. These post-retrieval IRs provide a uniform
analysis representation rather than historical compiler output; reviewers
therefore used source, IR, and relevant issue evidence when establishing
labels. Release checks enforce unique identifiers and complete source/IR
bundles; the archive documents the full schema and checks.

The final corpus distribution over the descriptive classes is shown in
Table~\ref{tab:taxonomy}; Section~\ref{sec:characterization} explains how the
classes are assigned and how they should be interpreted.

\begin{table*}[t]
\caption{Primary mechanism and boundary classes in the final corpus.}
\label{tab:taxonomy}
\centering
\small
\begin{tabularx}{\textwidth}{XrXr}
\toprule
\multicolumn{2}{c}{CISB mechanism} &
\multicolumn{2}{c}{Non-CISB boundary} \\
\midrule
Spatial/temporal memory safety & 44 & Equivalent or benign behavior & 86 \\
Data-state integrity & 41 & Source-authored unsafe behavior & 18 \\
Alias/provenance & 40 & No supported security consequence & 16 \\
Control flow & 29 & Incomplete/non-instantiating reduction & 16 \\
Runtime protection & 28 & Diagnostic-only/no runtime protection & 8 \\
Policy-dependent UB amplification & 25 & Compiler-process failure & 5 \\
ABI/target lowering & 22 & & \\
Indeterminate state & 18 & & \\
Concurrency/ordering & 10 & & \\
Sensitive-data handling & 8 & & \\
Resource/availability & 8 & & \\
Status-state integrity & 7 & & \\
\midrule
Total & 280 & Total & 149 \\
\bottomrule
\end{tabularx}
\end{table*}

\section{Manual Verification and Final Labels}
\label{sec:verification}

Mining produces candidate evidence, not ground truth. We therefore fixed the
429-row corpus before manual review and asked two undergraduate researchers
who had completed compiler coursework to independently classify every row.
Reviewers applied the three gates in Section~\ref{sec:definition}, inspecting
the C reduction, its standardized source--IR analysis artifacts, and public issue evidence when it
was needed to establish a historical target, patch, or manifestation.
They received a shared written codebook with worked anchor cases, but did not
conduct a formal calibration or training exercise before independent labeling.

Reviewers agreed on 369 of 429 rows (86.0\%) and disagreed on 60 (14.0\%).
The corresponding Cohen's $\kappa$ is 0.662. Of the disagreements, 52 had one
reviewer label the row CISB and the other non-CISB in one direction, and eight
in the reverse direction. The statistic is intentionally reported before
adjudication: the resolved labels are the benchmark target, not a measure of
independent reviewer agreement.

Each disagreement received a final evidence-based adjudication. The adjudicator
considered the competing classifications against G1--G3, the row's C
reduction, standardized IR bundle, and relevant issue context, then recorded a
final label and evidence-based rationale. The adjudicator was a doctoral
researcher in Computer Science. Adjudication is therefore not an
automatic majority vote. The resulting frozen target contains 280 CISBs
(65.3\%) and 149 non-CISB cases (34.7\%). The final data do
not present post-adjudication identity as inter-rater reliability; they provide
one frozen evaluation target for downstream methods.

This distinction matters because label noise and misclassification can distort
conclusions drawn from mined software data~\cite{herzig2013misclassification,croft2023dataquality}.
Retaining a fixed candidate set, pre-adjudication agreement, and a separate
final target makes the review protocol inspectable without implying that the
adjudicated labels are mechanically certain.

The disagreements reveal why a compiler-security benchmark cannot use issue
language alone. A row may omit the claimed trigger, show only a build-time
failure, inherit unsafe behavior from source, or have no security-bearing use
for an observed wrong value. Conversely, a compact reduction may preserve a
lost sanitizer check, invalid atomic behavior, or target-lowering error that a
generic wrong-code label does not capture. Final labels thus represent a
documented decision about the exact program, not a broad assertion about every
report in the issue family. The corpus contains 277 GCC-origin rows (166 CISB,
111 non-CISB) and 152 LLVM-origin rows (114 CISB, 38 non-CISB); these are
descriptive corpus counts, not estimates of compiler-defect prevalence.

\section{Mechanisms, Boundaries, and Characterization}
\label{sec:characterization}

\subsection{Primary taxonomy}

The binary target establishes whether a row is a CISB, but it does not explain
how a compiler action becomes security-relevant. We therefore assign every
CISB one dominant compiler-created mechanism and every non-CISB one primary
evidence boundary. A class captures the principal compiler-to-security chain,
not every possible downstream consequence. For example, a wrong stack-slot
access is classified as ABI/target lowering even when it could also affect
confidentiality or memory safety.

The taxonomy is derived after binary-label adjudication and is descriptive of
this corpus, rather than an independently validated labeling instrument.
Reviewers selected the same primary class on 252 of 429 rows (58.7\%;
Cohen's $\kappa=0.535$) after normalizing label wording and combining spatial
and temporal memory safety. The remaining 177 primary classes were resolved
through taxonomy adjudication using the row-level source, IR, and final-label
evidence. This distinction lets users treat the binary labels as the primary
benchmark target while using the taxonomy for error analysis and controlled
subset selection.

\subsection{What the corpus represents}

The most common CISB mechanisms are spatial/temporal memory safety (44 rows),
data-state integrity (41), and alias/provenance (40). This distribution shows
that CISBs are not confined to familiar dead-store-elimination examples. They
also arise when compilation changes object identity, lifetime state, a live
value flowing to a check, an ABI state, or a generated protection.

The negative set is equally informative. Equivalent or benign behavior is the
largest non-CISB boundary (86 rows), followed by source-authored unsafe
behavior (18), no supported security consequence (16), and incomplete or
non-instantiating reductions (16). These cases make a superficial classifier
look plausible while testing whether it matches the reviewed exact-row labels.
For instance, a removed operation is not necessarily a CISB if it is dead or
unobservable; a visible wrong value is not necessarily security-relevant if it
is printed and discarded; and a sanitizer report is not evidence that the
current reduction executes the required violation.

Compiler-origin counts characterize the released resource, not population
defect rates.

\section{Research Use and Reproducibility}
\label{sec:utility}

\subsection{Worked decision boundary}

The following pair operationalizes the exact-row decision boundary.
\texttt{cisb\_0068} is a CISB row from LLVM issue 129951. The historical report
uses Clang with \texttt{-O2 -Wall -fstrict-flex-arrays=3
-fsanitize=bounds}: it reports a direct \texttt{p->array[size]} access for
\texttt{size=3}, but omits the requested bounds check when the same
out-of-bounds address is formed as \texttt{\&p->array[size]} and then
dereferenced. This establishes G1 (incorrect sanitizer instrumentation), G2
(omission of requested runtime protection), and G3 (a concrete out-of-bounds
load executes without the requested detection). Its primary class is
\emph{runtime protection}.

By contrast, \texttt{cisb\_0148} also concerns a potentially unsafe value but
is a non-CISB row. Its source passes an uninitialized local integer to
\texttt{printf}; the associated GCC report concerns a missing
\texttt{-Wuninitialized} diagnostic. The unsafe read is source-authored, the
claimed compiler behavior is a warning rather than a changed runtime check, and
the exact row establishes no distinct compiler-created security consequence.
It consequently fails G1 and G2, and does not establish G3 for the CISB
definition. The non-CISB result is not an assertion that this program is safe;
it records only that the retained evidence does not support a CISB claim for
this exact row.

\subsection{Recommended evaluation and release use}

Construct an undirected graph connecting rows that share an issue identifier or
an identical source hash, and assign each connected component wholly to one
partition. Thus, related rows, including those connected transitively through
either relation, must not cross partitions. This matters because the 429 rows
have 283 issue IDs and 366 hashes; 88 issue IDs recur across 234 rows and 62
hashes recur across 125 rows. Taxonomy and adjudication rationale are reserved
for post-hoc analysis and must not be detector inputs. Report-text retrieval
should be declared separately from a method that inspects only released C and
IR.

Because the corpus has 280 CISBs and 149 deliberately difficult non-CISBs,
aggregate accuracy is insufficient. We recommend CISB precision, recall, F1,
a confusion matrix, and post-hoc errors by mechanism and boundary class.
Analyzer studies should also state compiler version, target, and optimization
level: contemporary non-reproduction of a historically supported case is not,
by itself, a label error.

The public archive provides the canonical CSV, per-sample C/IR bundles, README,
and construction materials at
\url{https://doi.org/10.6084/m9.figshare.31797592}. It excludes private review
material, credentials, and local paths. The dataset is not a census: it is
restricted to retainable C reductions from GCC and LLVM reports and may
underrepresent other ecosystems, historical environments, and failures without
compact reproducers.

\section{Related Work and Conclusion}
\label{sec:related}

Compiler fuzzing, differential testing, and translation validation aim to find
or rule out semantic compiler errors~\cite{chen2020compilertestingsurvey,yang2011csmith,le2014emi,lopes2021alive2}.
Empirical compiler-bug studies characterize the reports and optimization
defects in GCC and LLVM~\cite{sun2016compilerbugs,zhou2021optimization}.
These efforts motivate our data sources, but their targets are broader than the
security consequence of an exact C reduction paired with standardized IR.
Secure-compilation research
and work on the correctness--security gap show why semantic preservation alone
does not guarantee a selected security property~\cite{dsilva2015gap,deng2018securing,patrignani2019securecompilation}.
Xu et al.~\cite{xu2023silentbugs} provide the closest empirical CISB study;
\bench complements it with an auditable benchmark target, hard negatives, and
paired optimization-level artifacts. Formal compiler correctness and
constant-time-preservation work provide complementary guarantees for selected
compiler or security properties~\cite{leroy2009compcert,barthe2020ctcompcert},
while binary security analysis can check properties such as constant time and
secret erasure at a different layer~\cite{daniel2022binsecrel}.

Application-vulnerability datasets such as Big-Vul and MegaVul support learning
from CVE-linked source changes~\cite{fan2020bigvul,ni2024megavul}; related
resources include CVEfixes~\cite{bhandari2021cvefixes}.
They do not,
however, provide compiler-generated IR artifacts or label the compiler's causal
role. Recent MSR data papers similarly emphasize reproducible security
instances, quality-controlled labels, or manually validated task data
\cite{ji2026rustxec,lu2025icvul,kahapola2026javabackports}.

\bench fills this resource gap with 429 inspected C reductions and
standardized IR bundles, 280 final CISBs, 149 hard non-CISB cases, and
descriptive mechanism/boundary annotations.
The task evaluates prediction of the reviewed exact-row label from released C
and IR. A prediction does not itself establish the historical
compiler-to-security mechanism, which may additionally require issue evidence
and the affected compiler configuration.

\newpage
\bibliographystyle{IEEEtran}
\bibliography{bib/references}

@inproceedings{yang2011csmith,
  author    = {Yang, Xuejun and Chen, Yang and Eide, Eric and Regehr, John},
  title     = {Finding and Understanding Bugs in {C} Compilers},
  booktitle = {Proceedings of the 32nd ACM SIGPLAN Conference on Programming Language Design and Implementation},
  pages     = {283--294},
  year      = {2011},
  doi       = {10.1145/1993498.1993532}
}

@inproceedings{le2014emi,
  author    = {Le, Vu and Afshari, Mehrdad and Su, Zhendong},
  title     = {Compiler Validation via Equivalence Modulo Inputs},
  booktitle = {Proceedings of the 35th ACM SIGPLAN Conference on Programming Language Design and Implementation},
  pages     = {216--226},
  year      = {2014},
  doi       = {10.1145/2594291.2594334}
}

@inproceedings{sun2016compilerbugs,
  author    = {Sun, Chengnian and Le, Vu and Zhang, Qirun and Su, Zhendong},
  title     = {Toward Understanding Compiler Bugs in {GCC} and {LLVM}},
  booktitle = {Proceedings of the 25th International Symposium on Software Testing and Analysis},
  pages     = {294--305},
  year      = {2016},
  doi       = {10.1145/2931037.2931074}
}

@inproceedings{dsilva2015gap,
  author    = {D'Silva, Vijay and Payer, Mathias and Song, Dawn},
  title     = {The Correctness-Security Gap in Compiler Optimization},
  booktitle = {2015 IEEE Security and Privacy Workshops},
  pages     = {73--87},
  year      = {2015},
  doi       = {10.1109/SPW.2015.33}
}

@inproceedings{xu2023silentbugs,
  author    = {Xu, Jianhao and Lu, Kangjie and Du, Zhengjie and Ding, Zhu and Li, Linke and Wu, Qiushi and Payer, Mathias and Mao, Bing},
  title     = {Silent Bugs Matter: A Study of Compiler-Introduced Security Bugs},
  booktitle = {32nd USENIX Security Symposium (USENIX Security 23)},
  pages     = {3655--3672},
  year      = {2023},
  url       = {https://www.usenix.org/conference/usenixsecurity23/presentation/xu-jianhao}
}

@article{zhou2021optimization,
  author  = {Zhou, Zhide and Ren, Zhilei and Gao, Guojun and Jiang, He},
  title   = {An Empirical Study of Optimization Bugs in {GCC} and {LLVM}},
  journal = {Journal of Systems and Software},
  volume  = {174},
  pages   = {110884},
  year    = {2021},
  doi     = {10.1016/j.jss.2020.110884}
}

@inproceedings{bui2022vul4j,
  author    = {Bui, Quang-Cuong and Scandariato, Riccardo and Ferreyra, Nicol{\'a}s E. D{\'i}az},
  title     = {{Vul4J}: A Dataset of Reproducible Java Vulnerabilities Geared Towards the Study of Program Repair Techniques},
  booktitle = {Proceedings of the 19th International Conference on Mining Software Repositories},
  pages     = {464--468},
  year      = {2022},
  doi       = {10.1145/3524842.3528482}
}

@inproceedings{wang2012undefined,
  author    = {Wang, Xi and Chen, Haogang and Cheung, Alvin and Jia, Zhihao and Zeldovich, Nickolai and Kaashoek, M. Frans},
  title     = {Undefined Behavior: What Happened to My Code?},
  booktitle = {Proceedings of the 3rd Asia-Pacific Workshop on Systems},
  year      = {2012}
}

@article{deng2018securing,
  author  = {Deng, Chaoqiang and Namjoshi, Kedar S.},
  title   = {Securing a Compiler Transformation},
  journal = {Formal Methods in System Design},
  volume  = {53},
  number  = {2},
  pages   = {166--188},
  year    = {2018},
  doi     = {10.1007/s10703-017-0313-8}
}

@inproceedings{bhandari2021cvefixes,
  author    = {Bhandari, Guru Prasad and Naseer, Amara and Moonen, Leon},
  title     = {{CVEfixes}: Automated Collection of Vulnerabilities and Their Fixes from Open-Source Software},
  booktitle = {Proceedings of the 17th International Conference on Predictive Models and Data Analytics in Software Engineering},
  pages     = {30--39},
  year      = {2021}
}

@inproceedings{ni2024megavul,
  author    = {Ni, Chao and Shen, Liyu and Yang, Xiaohu and Zhu, Yan and Wang, Shaohua},
  title     = {{MegaVul}: A {C/C++} Vulnerability Dataset with Comprehensive Code Representations},
  booktitle = {Proceedings of the 21st International Conference on Mining Software Repositories},
  pages     = {738--742},
  year      = {2024},
  doi       = {10.1145/3643991.3644886}
}

@inproceedings{wang2013stack,
  author    = {Wang, Xi and Zeldovich, Nickolai and Kaashoek, M. Frans and Solar-Lezama, Armando},
  title     = {Towards Optimization-Safe Systems: Analyzing the Impact of Undefined Behavior},
  booktitle = {Proceedings of the 24th ACM Symposium on Operating Systems Principles (SOSP '13)},
  year      = {2013},
  pages     = {260--275},
  publisher = {ACM},
  doi       = {10.1145/2517349.2522728}
}

@article{lee2017taming,
  author    = {Lee, Juneyoung and Kim, Yoonseung and Song, Youngju and Hur, Chung-Kil and Das, Sanjoy and Majnemer, David and Regehr, John and Lopes, Nuno P.},
  title     = {Taming Undefined Behavior in {LLVM}},
  journal   = {ACM SIGPLAN Notices},
  volume    = {52},
  number    = {6},
  year      = {2017},
  pages     = {633--647},
  note      = {Proceedings of PLDI 2017},
  doi       = {10.1145/3062341.3062343}
}

@inproceedings{yang2017dse,
  author    = {Yang, Zhaomo and Johannesmeyer, Brian and Olesen, Anders Trier and Lerner, Sorin and Levchenko, Kirill},
  title     = {Dead Store Elimination (Still) Considered Harmful},
  booktitle = {Proceedings of the 26th USENIX Security Symposium (USENIX Security '17)},
  year      = {2017},
  pages     = {1025--1040},
  publisher = {USENIX Association}
}

@inproceedings{simon2018whatyouc,
  author    = {Simon, Laurent and Chisnall, David and Anderson, Ross},
  title     = {What You Get is What You {C}: Controlling Side Effects in Mainstream {C} Compilers},
  booktitle = {2018 IEEE European Symposium on Security and Privacy (EuroS\&P)},
  year      = {2018},
  pages     = {1--15},
  publisher = {IEEE},
  doi       = {10.1109/EuroSP.2018.00009}
}

@article{isemann2023dontlookub,
  author    = {Isemann, Raphael and Giuffrida, Cristiano and Bos, Herbert and van der Kouwe, Erik and von Gleissenthall, Klaus},
  title     = {Don't Look {UB}: Exposing Sanitizer-Eliding Compiler Optimizations},
  journal   = {Proceedings of the ACM on Programming Languages},
  volume    = {7},
  number    = {PLDI},
  articleno = {143},
  year      = {2023},
  doi       = {10.1145/3591257}
}

@inproceedings{xu2023warpattack,
  author    = {Xu, Jianhao and Di Bartolomeo, Luca and Toffalini, Flavio and Mao, Bing and Payer, Mathias},
  title     = {{WarpAttack}: Bypassing {CFI} through Compiler-Introduced Double-Fetches},
  booktitle = {2023 IEEE Symposium on Security and Privacy (S\&P)},
  year      = {2023},
  pages     = {1271--1288},
  publisher = {IEEE},
  doi       = {10.1109/SP46215.2023.10179433}
}

@article{barthe2020ctcompcert,
  author    = {Barthe, Gilles and Blazy, Sandrine and Gr{\'e}goire, Benjamin and Hutin, R{\'e}mi and Laporte, Vincent and Pichardie, David and Trieu, Alix},
  title     = {Formal Verification of a Constant-Time Preserving {C} Compiler},
  journal   = {Proceedings of the ACM on Programming Languages},
  volume    = {4},
  number    = {POPL},
  articleno = {7},
  year      = {2020},
  doi       = {10.1145/3371075}
}

@article{daniel2022binsecrel,
  author    = {Daniel, Lesly-Ann and Bardin, S{\'e}bastien and Rezk, Tamara},
  title     = {{Binsec/Rel}: Symbolic Binary Analyzer for Security with Applications to Constant-Time and Secret-Erasure},
  journal   = {ACM Transactions on Privacy and Security},
  volume    = {26},
  number    = {2},
  articleno = {11},
  year      = {2022},
  doi       = {10.1145/3563037}
}

@inproceedings{schneider2025breakingbad,
  author    = {Schneider, Moritz and Lain, Daniele and Puddu, Ivan and Dutly, Nicolas and Capkun, Srdjan},
  title     = {Breaking Bad: How Compilers Break Constant-Time Implementations},
  booktitle = {Proceedings of the 20th ACM Asia Conference on Computer and Communications Security (ASIA CCS '25)},
  year      = {2025},
  publisher = {ACM},
  doi       = {10.1145/3708821.3733909}
}

@inproceedings{kaufmann2016constanttime,
  author    = {Kaufmann, Thierry and Pelletier, Herv{\'e} and Vaudenay, Serge and Villegas, Karine},
  title     = {When Constant-Time Source Yields Variable-Time Binary: Exploiting {Curve25519-donna} Built with {MSVC} 2015},
  booktitle = {Cryptology and Network Security (CANS 2016)},
  series    = {Lecture Notes in Computer Science},
  volume    = {10052},
  year      = {2016},
  pages     = {573--582},
  publisher = {Springer},
  doi       = {10.1007/978-3-319-48965-0_36}
}

@article{patrignani2019securecompilation,
  author    = {Patrignani, Marco and Ahmed, Amal and Clarke, Dave},
  title     = {Formal Approaches to Secure Compilation: A Survey of Fully Abstract Compilation and Related Work},
  journal   = {ACM Computing Surveys},
  volume    = {51},
  number    = {6},
  articleno = {125},
  year      = {2019},
  doi       = {10.1145/3280984}
}

@article{chen2020compilertestingsurvey,
  author    = {Chen, Junjie and Patra, Jibesh and Pradel, Michael and Xiong, Yingfei and Zhang, Hongyu and Hao, Dan and Zhang, Lu},
  title     = {A Survey of Compiler Testing},
  journal   = {ACM Computing Surveys},
  volume    = {53},
  number    = {1},
  articleno = {4},
  year      = {2020},
  doi       = {10.1145/3363562}
}

@inproceedings{regehr2012reduction,
  author    = {Regehr, John and Chen, Yang and Cuoq, Pascal and Eide, Eric and Ellison, Chucky and Yang, Xuejun},
  title     = {Test-Case Reduction for {C} Compiler Bugs},
  booktitle = {Proceedings of the 33rd ACM SIGPLAN Conference on Programming Language Design and Implementation (PLDI '12)},
  year      = {2012},
  pages     = {335--346},
  publisher = {ACM},
  doi       = {10.1145/2254064.2254104}
}

@article{lecoeur2023reconditioning,
  author    = {Lecoeur, Bastien and Mohsin, Hasan and Donaldson, Alastair F.},
  title     = {Program Reconditioning: Avoiding Undefined Behaviour When Finding and Reducing Compiler Bugs},
  journal   = {Proceedings of the ACM on Programming Languages},
  volume    = {7},
  number    = {PLDI},
  articleno = {180},
  year      = {2023},
  doi       = {10.1145/3591294}
}

@inproceedings{chen2013taming,
  author    = {Chen, Yang and Groce, Alex and Zhang, Chaoqiang and Wong, Weng-Keen and Fern, Xiaoli and Eide, Eric and Regehr, John},
  title     = {Taming Compiler Fuzzers},
  booktitle = {Proceedings of the 34th ACM SIGPLAN Conference on Programming Language Design and Implementation (PLDI '13)},
  year      = {2013},
  pages     = {197--208},
  publisher = {ACM},
  doi       = {10.1145/2491956.2462173}
}

@inproceedings{yang2023dedup,
  author    = {Yang, Chen and Chen, Junjie and Fan, Xingyu and Jiang, Jiajun and Sun, Jun},
  title     = {Silent Compiler Bug De-duplication via Three-Dimensional Analysis},
  booktitle = {Proceedings of the 32nd ACM SIGSOFT International Symposium on Software Testing and Analysis (ISSTA '23)},
  year      = {2023},
  pages     = {677--689},
  publisher = {ACM},
  doi       = {10.1145/3597926.3598087}
}

@article{marcozzi2019matter,
  author    = {Marcozzi, Micha{\"e}l and Tang, Qiyi and Donaldson, Alastair F. and Cadar, Cristian},
  title     = {Compiler Fuzzing: How Much Does It Matter?},
  journal   = {Proceedings of the ACM on Programming Languages},
  volume    = {3},
  number    = {OOPSLA},
  articleno = {155},
  year      = {2019},
  doi       = {10.1145/3360581}
}

@article{leroy2009compcert,
  author    = {Leroy, Xavier},
  title     = {Formal Verification of a Realistic Compiler},
  journal   = {Communications of the ACM},
  volume    = {52},
  number    = {7},
  year      = {2009},
  pages     = {107--115},
  doi       = {10.1145/1538788.1538814}
}

@article{lopes2021alive2,
  author    = {Lopes, Nuno P. and Lee, Juneyoung and Hur, Chung-Kil and Liu, Zhengyang and Regehr, John},
  title     = {{Alive2}: Bounded Translation Validation for {LLVM}},
  booktitle = {Proceedings of the 42nd ACM SIGPLAN International Conference on Programming Language Design and Implementation (PLDI '21)},
  year      = {2021},
  pages     = {65--79},
  publisher = {ACM},
  doi       = {10.1145/3453483.3454030}
}

@inproceedings{herzig2013misclassification,
  author    = {Herzig, Kim and Just, Sascha and Zeller, Andreas},
  title     = {It's Not a Bug, It's a Feature: How Misclassification Impacts Bug Prediction},
  booktitle = {Proceedings of the 35th International Conference on Software Engineering (ICSE '13)},
  year      = {2013},
  pages     = {392--401},
  publisher = {IEEE},
  doi       = {10.1109/ICSE.2013.6606585}
}

@inproceedings{croft2023dataquality,
  author    = {Croft, Roland and Babar, M. Ali and Kholoosi, M. Mehdi},
  title     = {Data Quality for Software Vulnerability Datasets},
  booktitle = {Proceedings of the 45th IEEE/ACM International Conference on Software Engineering (ICSE '23)},
  year      = {2023},
  pages     = {121--133},
  publisher = {IEEE},
  doi       = {10.1109/ICSE48619.2023.00022}
}

@inproceedings{fan2020bigvul,
  author    = {Fan, Jiahao and Li, Yi and Wang, Shaohua and Nguyen, Tien N.},
  title     = {A {C/C++} Code Vulnerability Dataset with Code Changes and {CVE} Summaries},
  booktitle = {Proceedings of the 17th International Conference on Mining Software Repositories (MSR '20)},
  year      = {2020},
  pages     = {508--512},
  publisher = {ACM},
  doi       = {10.1145/3379597.3387501}
}

@inproceedings{tomassi2019bugswarm,
  author    = {Tomassi, David A. and Dmeiri, Naji and Wang, Yichen and Bhowmick, Antara and Liu, Yen-Chuan and Devanbu, Premkumar T. and Vasilescu, Bogdan and Rubio-Gonz{\'a}lez, Cindy},
  title     = {{BugSwarm}: Mining and Continuously Growing a Dataset of Reproducible Failures and Fixes},
  booktitle = {Proceedings of the 41st International Conference on Software Engineering (ICSE '19)},
  year      = {2019},
  pages     = {339--349},
  publisher = {IEEE},
  doi       = {10.1109/ICSE.2019.00048}
}

@inproceedings{silva2024gitbugjava,
  author    = {Silva, Andr{\'e} and Saavedra, Nuno and Monperrus, Martin},
  title     = {{GitBug-Java}: A Reproducible Benchmark of Recent {Java} Bugs},
  booktitle = {Proceedings of the 21st International Conference on Mining Software Repositories (MSR '24)},
  year      = {2024},
  pages     = {118--122},
  publisher = {ACM},
  doi       = {10.1145/3643991.3644884}
}

@inproceedings{lu2025icvul,
  author    = {Lu, Chaomeng and Li, Tianyu and Dehaene, Toon and Lagaisse, Bert},
  title     = {{ICVul}: A Well-Labeled {C/C++} Vulnerability Dataset with Comprehensive Metadata and {VCCs}},
  booktitle = {Proceedings of the 22nd IEEE/ACM International Conference on Mining Software Repositories},
  pages     = {154--158},
  year      = {2025},
  doi       = {10.1109/MSR66628.2025.00034}
}

@inproceedings{ji2026rustxec,
  author    = {Ji, Zhengjie and Wang, Xin and Wang, Lingxiang and Li, Geng and Yang, Fan and Zhang, Ying},
  title     = {{RustXec}: A Vulnerability Reproduction Dataset for Assessing Security Risks in Open-Source {Rust} Applications},
  booktitle = {Proceedings of the 23rd International Conference on Mining Software Repositories},
  pages     = {590--594},
  year      = {2026},
  doi       = {10.1145/3793302.3793307}
}

@inproceedings{kahapola2026javabackports,
  author    = {Kahapola, Kaushal and Galappaththi, Sharada and Ranasinghe, Dinith and Shariffdeen, Ridwan Salihin and de Silva, Nisansa and Perera, Srinath and Wickramanayake, Sandareka},
  title     = {{JavaBackports}: A Dataset for Benchmarking Automated Backporting in {Java}},
  booktitle = {Proceedings of the 23rd International Conference on Mining Software Repositories},
  pages     = {712--716},
  year      = {2026},
  doi       = {10.1145/3793302.3793331}
}
\end{document}